\documentclass[journal=acsnano,manuscript=article]{achemso}

\usepackage{physics}
\usepackage[version=3]{mhchem} 
\usepackage{amssymb,amsmath}
\usepackage{float}
\usepackage{units}
\usepackage[dvipsnames]{xcolor}

\newcommand{\ears}{{\it Active Excitonic Regions\,}}
\newcommand{\oscreen}{{\it Over Screening}\,}
\newcommand{\paulib}{{\it Pauli Blocking}\,}

\def\kk{{\mathbf k}}

\newcommand{\inlinecite}[1]{(\!\!\citenum{#1})}
\author{Valerie Smejkal}
\email{valerie.smejkal@tuwien.ac.at}
\author{Florian Libisch}
\email{florian.libisch@tuwien.ac.at}
\affiliation[TU Wien]
{Vienna University of Technology, Institute for Theoretical Physics, 1040 Vienna, Austria, EU}
\author{Alejandro Molina-Sanchez}
\affiliation[CNR]
{Institute of Materials Science (ICMUV), University of Valencia, Valencia, Spain, EU}
\author{Ludger Wirtz}
\affiliation[UniLu]
{Department of Physics and Materials Science, University of Luxembourg, 1511 Luxembourg, Luxembourg, EU}
\author{Andrea Marini}
\affiliation[CNR]
{CNR-ISM, Division of Ultrafast Processes in Materials (FLASHit), Area della Ricerca di Roma 1, Via Salaria Km 29.3, I-00016 Monterotondo Scalo, Italy, EU}

\title
[Time-dependent screening explains the ultrafast excitonic signal rise in 2D semiconductors]
{Time-dependent screening explains the ultrafast excitonic signal rise in 2D semiconductors}

\abbreviations{BSE,COHSEX,BGR,EBR,PB}

\renewcommand{\vec}[1]{\mathbf{ #1 }}
\begin{document}


\begin{abstract}
We calculate the time evolution of the transient reflection signal in an MoS$_2$ monolayer on a SiO$_2$/Si substrate using first-principles out-of-equilibrium
real-time methods. Our simulations provide a simple and intuitive physical picture for the delayed, yet ultrafast, evolution of the signal whose rise time depends on the
excess energy of the pump laser: at laser energies above the A- and B-exciton, the pump pulse excites electrons and holes far away from the K valleys in the
first Brillouin zone. Electron-phonon and hole-phonon scattering lead to a gradual relaxation of the carriers towards small {\it Active Excitonic Regions}
around K, enhancing the dielectric screening. 
The accompanying time-dependent band gap renormalization dominates over Pauli blocking and the excitonic binding energy renormalization. 
This explains the delayed buildup of the transient reflection signal of the probe pulse, in excellent agreement with recent experimental data. 
Our results show that the observed delay is not a unique signature of an exciton formation
process but rather caused by coordinated carrier dynamics and its influence on the screening.
\end{abstract}

The semiconducting monolayer transition metal dichalcogenides\,(TMDs) have been subject to intense research, not only because of their potential applications in
electronic devices, but also for their exciting optical properties \cite{Srivastava2015c,Berkelbach2018,Mueller2018}. When reducing the number of layers from
bulk to a monolayer, TMDs transition
from an indirect band gap to a direct band gap semiconductor \cite{Splendiani2010,Mak2010}, opening the path for novel optical applications. One of their most
intriguing
properties is the presence of strongly bound excitons that are observable at room temperature with binding energies of hundreds of meV \cite{Qiu2013,Molina-Sanchez2013,Park2018a,Hill2015}.
While the equilibrium properties of these excitons have been extensively studied, the associated dynamics are subject to an intense debate. Indeed,
pump-probe experiments allow to investigate the very initial states of the photoexcited carriers posing a fundamental question: do the
photo-excited electron-hole pairs bind instantaneously? And if they bind dynamically, on what time-scale does this binding occur?

After a pioneering study in 2003 describing the formation time of a quasi--particle in silicon~\cite{Hase2003/11/06/printa} there has been an explosive quest in the last years for a similar investigation of the real-time formation of excitons. Several studies have tried to pinpoint the time it takes to build up an
excitonic population among the photo-excited electron-hole pairs by means of mainly two different experimental techniques: transient absorption\,(TR--ABS) and
transient Time--Resolved-Angle--Resolved-Photoemission\, (TR--ARPES). In the case of TR--ABS~\cite{Ceballos2016,Trovatello2020g,Steinleitner2017} the change in the optical absorption of the probe is monitored as a function of the delay and energy of the pump field. These works connect the exciton formation time to the rise/decay time of the transient absorption/reflection signal, yet find a wide range of time scales from 30\,fs to 300\,fs.
By contrast, TR--ARPES~\cite{mado2020directly,Dendzik2020,Steinhoff2017} connects the observation of the exciton formation to the appearance of additional peaks in the ARPES spectrum which cannot be explained in terms of an independent particle approach.

By construction, the approach based on TR--ABS is more more difficult to interpret compared to the TR--ARPES to trace the peculiar signature
of an excitonic state formation: in TR--ARPES there exists an energy region below the absorption onset that, in the equilibrium condition, is forbidden to single--particle states. Any state appearing in this region, such as an excitonic state, is therefore most probably not a single-particle state.
Conversely, TR--ABS in general monitors the delay-dependent change in the probe absorption, transmission or reflection. The resulting pump-excited
excitonic states overlap with standard probe absorption peaks, which are also influenced by the dynamics. This makes the analysis more delicate and indirect.

\begin{figure}[H]
	\centering
	\includegraphics[width=7in]{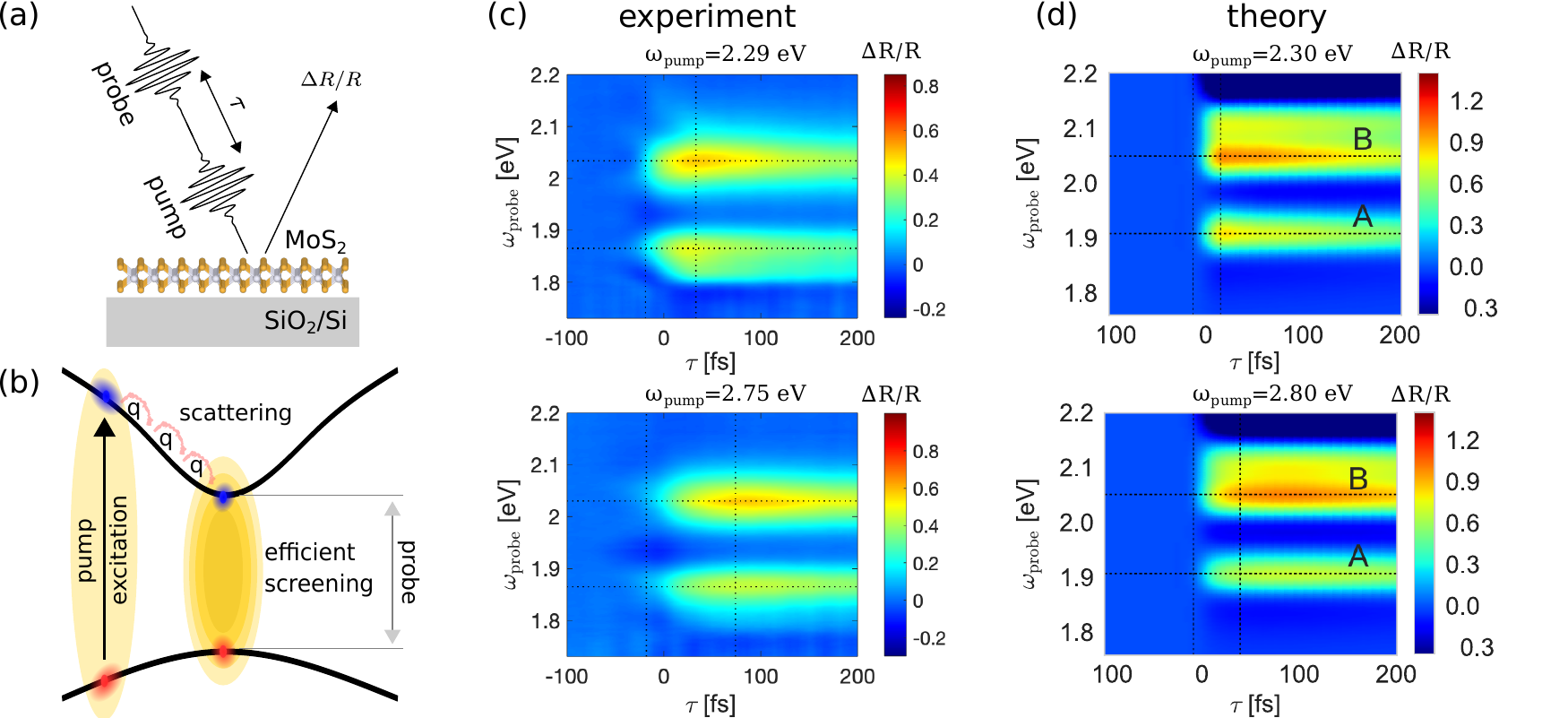}
	\caption{(a) Sketch of the pump--probe experimental setup in reflection geometry. (b) Schematic representation of the photo--induced dynamics: electron--hole pairs are excited by the pump pulse. As electrons and holes scatter to the band extrema, their effect on the screening becomes more efficient leading to three effects: Pauli Blocking, the renormalization of the band gap and the renormalization of the excitonic binding energy. (c) Experimental (adapted from Ref.~\inlinecite{Trovatello2020g}) and (d) theoretical energy- and delay-time-dependent transient reflectivity maps for two different pump energies $\omega_\mathrm{pump}$. The vertical dashed lines denote 10\% and 90\% of the maximum signal.}
	\label{fig:schematics}
\end{figure}

The most recent TR--ABS experiment~\cite{Trovatello2020g} aimed at observing the formation of excitons investigates a monolayer of MoS$_2$ excited by a short pump pulse with energies $\omega_\mathrm{pump}$ tunable between the A-excitonic resonance at 1.90\,eV up to energies ($\approx2.80$\,eV) far in the electron-hole continuum [see Fig.~\ref{fig:schematics}(a)]. A broadband probe pulse, covering the spectral extent between 1.75\,eV and 2.4\,eV, samples the evolution of the A- and B-exciton and the resulting change in the reflectivity. The excellent resolution of the setup
($\sim 30$\,fs) allows to monitor the dynamics within the first hundred femtoseconds: as the pump energy increases, there is a delay in the rise time of the signal of the order of $\leq 35$\,fs. 

The experimental time- and energy-dependent change in the probe reflectivity $\Delta R/R$ at different pump energies 2.29\,eV and 2.70\,eV [Fig.~\ref{fig:schematics}(c)] clearly shows an increase of the delay between the pump and the probe maximum $\Delta R/R$. The authors attribute this delayed rise time to the buildup of an incoherent exciton population mediated by phonons~\cite{Trovatello2020g}. This interpretation is based on the Bloch equations for the excitonic amplitude and occupation~\cite{Katsch2018}. The idea is that the photo-excited carriers form a correlated electron-hole pair immediately after the pump and all the physics of the following dynamics is interpreted in terms of excitonic scattering events.

While a description based on Bloch equations following Ref.~\inlinecite{Katsch2018} certainly covers important aspects of the physics involved, it ignores the change in
screening by excited carriers leading to band gap renormalization\,(BGR) and an excitonic binding energy renormalization\,(EBR) which has been shown to be among
the driving forces in equilibrium as well as non-equilibrium experiments of monolayer TMDs~\cite{Wood2020,Gao2017,Cunningham2017,Zhao2020,Pogna2016,Yao2017b}.
In this work we propose an alternative interpretation of the pump-energy-dependent rise time of the probe $\Delta R/R$ based on the detailed dynamics of the carrier distribution in the MoS$_2$ Brillouin zone (BZ) driven by electron-phonon scattering. We identify specific regions around the K edges, that we refer to
as \ears. These \ears are closely connected to the $\mathbf{k}$-space distribution of the A- and B-excitons. We demonstrate that carrier migration from and to the \ears modifies both the excitonic packet and the dielectric screening of the material. The latter renormalizes both the band gap (BGR) and the excitonic binding energy (EBR).

Comparing our simulated energy and delay-time-dependent transient reflectivity maps $\Delta R(\omega_{\mathrm{probe}},\Delta\tau)$ [Fig.~\ref{fig:schematics}(d)] to the measurement~\cite{Trovatello2020} [Fig.~\ref{fig:schematics}(c)], we find excellent agreement. More importantly, our fully ab--initio approach implemented in the \texttt{Yambo} code~\cite{Sangalli2019b,Marini2009a} avoids any adjustable parameters.

In our simulations, we accurately reproduce the experimental setup of Ref.\inlinecite{Trovatello2020g}:
we simulate a monolayer of MoS$_2$ deposited on a SiO$_2$/Si substrate using {\it ab initio} atomistic methods for the crystal and the Fresnel equations~\cite{Byrnes2016} to describe photonic effects from the substrate. The electronic levels, phonons and electron-phonon scattering potentials are calculated within
Density--Functional Theory~\cite{Giannozzi2020}. The out--of--equilibrium dynamics is performed with the \texttt{Yambo} code in the Kohn--Sham basis. The
theoretical description in \texttt{Yambo} is based on Non--Equilibrium Green's Function Theory\,(NEGF)~\cite{Marini2013}.
The implementation of the NEGF in \texttt{Yambo} has been used successfully in numerous applications to explain the non-equilibrium dynamics in photoexcited
semiconductors~\cite{Attaccalite2011,Marini2013,Sangalli2015,Pogna2016,Sangalli2016,Molina-Sanchez2017,Wang2018,Roth2019,Trovatello2020}.
More details are given in the Supplementary Information (SI).

We also adopt the same pump-probe geometry of Ref.\inlinecite{Trovatello2020g}:
the MoS$_2$/SiO$_2$/Si stack is pumped by a linearly polarized Gaussian-envelope laser pulse of a full-width half-maximum (of the intensity) of $20~\mathrm{fs}$ and a
peak intensity of $I_0=5\cdot 10^4~\mathrm{W/cm}^2$ inside the material. By including the electron--hole interaction, we correctly describe the transition from resonant to off--resonant excitations of the A- and B-excitons~\cite{Li2014,Molina-Sanchez2015,Molina-Sanchez2016}, as the pump pulse energy
increases from the A-exciton energy ($\omega_\mathrm{pump}=1.90$\,eV) to the continuum ($\omega_\mathrm{pump}=3.20$\,eV).
To simulate the transient reflection of the probe laser, we use the carrier distribution at delay time $\tau$ to calculate the time-dependent static screening
$\varepsilon_s(\vec{q},\tau)$, the pump-induced
shift of the energy levels and the change in the optical properties from the Time--Dependent Bethe--Salpeter Equation\,(BSE)~\cite{Sangalli2016,Perfetto2016},
see Methods. This approach leads to a consistent description of the carrier dynamics and of the effects induced on the electronic and excitonic
states: BGR, EBR and the excitonic Pauli Blocking\,(PB).

\section{Results and Discussion}

\begin{figure}[h]
\includegraphics[width=3.25in]{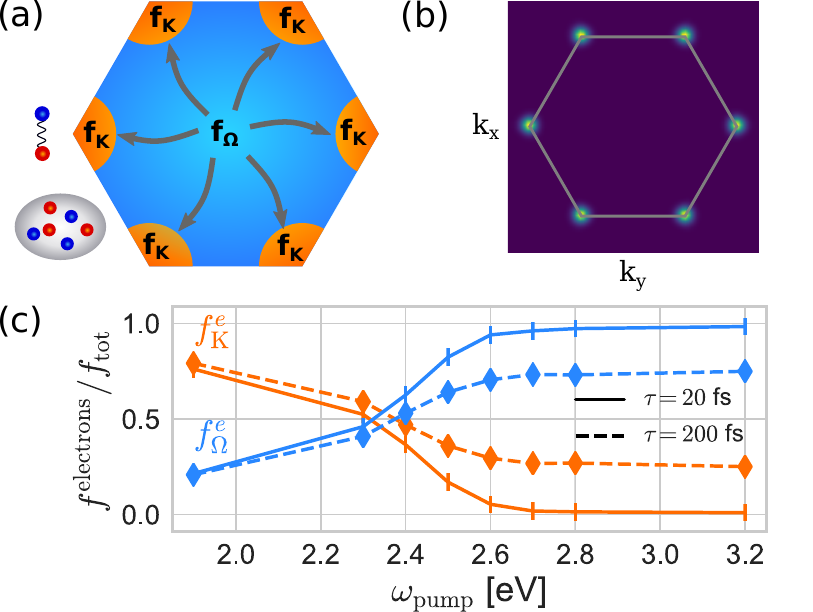}
\caption{Pump-energy-dependent time--evolution of the carriers $\kk$-space distribution following the initial photo--excitation. 
(a) Splitting of the Brillouin zone into six \ears around the band extrema at the $\mathrm{K}$ points with carrier occupation  $f_\mathrm{K}$. The radius
of those regions is around 7\% of the BZ length. The
rest of the Brillouin zone is labeled as $\Omega$ and has carrier occupation $f_\Omega$. 
(b) Reciprocal space distribution of the A-exciton. The excitonic distribution defines the \ears.
(c) Pump-energy-dependent contribution of $f_\mathrm{K}$ and of $f_\Omega$ to the total number of excited electrons right after the
pump (solid line) and at a 200\,fs delay time (dashed line).}
\label{fig:carriers}
\end{figure}

In order to closely connect the photo--excitation geometry to the microscopic carrier dynamics, we partition the hexagonal BZ of monolayer MoS$_2$ in \ears around the $\mathrm{K}$ points\,[see Fig.~\ref{fig:carriers}\,(a)]. These regions coincide with the fraction of the BZ where the A- and B-excitons are mostly distributed, as clear from Fig.~\ref{fig:carriers}\,(b). 
The occupations inside and outside the \ears are defined as $f_\mathrm{K}$ and $f_\Omega$, respectively.

The dependence of $f_\mathrm{K}$ and $f_\Omega$ on the pump energy and delay time [Fig.~\ref{fig:carriers}\,(c)] conveys a simple and clear 
message: during the photo-excitation, the carriers are injected into electron-hole pairs that correspond to the exciton resonant with the pump laser.
As the pump energy increases above the A- and B-exciton, the carriers are pumped into regions outside the \ears.
After this initial population, the subsequent electron dynamics leads to a gradual re--population of the \ears.

The physics of this experiment and the consequent change and rise of the transient absorption signal can naturally be traced back to the phenomena that are
activated by the electrons moving into and out of the \ears: \paulib\, and \oscreen. \paulib\, occurs whenever the carriers scatter into a state occupied by the
exciton, i.e.\,into the \ears. The enhancement of the screening properties\,(\oscreen) also depends on $f_\mathrm{K}$ and $f_\Omega$, even if the basic mechanism is different as discussed below. It is the  \oscreen\, which leads to the BGR and to the EBR.

\begin{figure}[ht]
\includegraphics[width=3.25in]{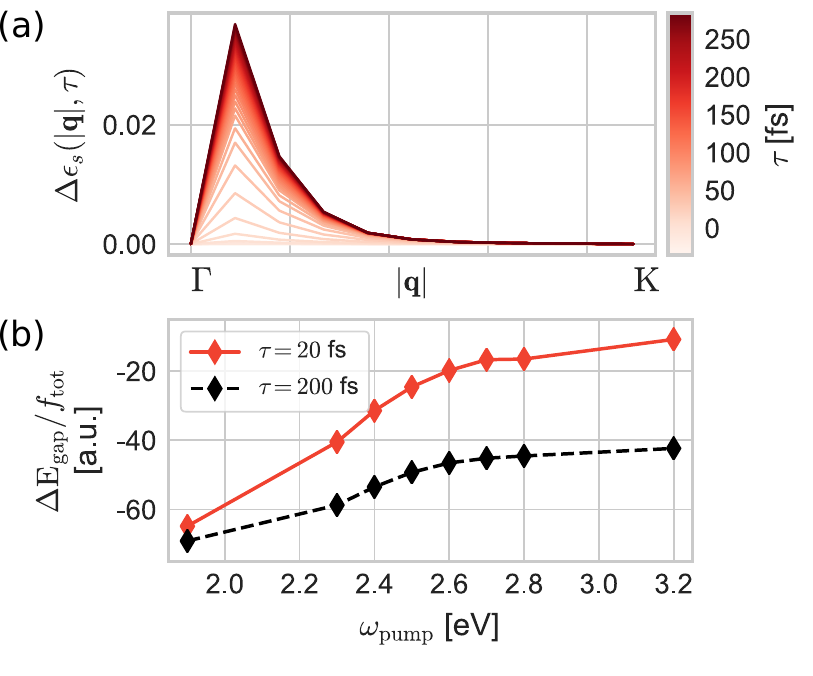}
\caption{(a) Carrier driven change of the static screening $\Delta \varepsilon_s(\vec{q},\tau)$ for $\omega_\mathrm{pump}=2.50$\,eV for different times $\tau$. As the carriers
relax to the \ears, the screening increases (\oscreen). (b) Pump-energy- and time-dependent evolution of the average band gap directly after the pulse (solid red line) and
after 200\,fs (dashed black line). Different pump energies are normalized to the total number of excited carriers $f_\mathrm{tot}(\omega_\mathrm{pump})$ which
varies due to the energy-dependent optical absorption cross section.}
\label{fig:screening}
\end{figure}

While in the case of \paulib the simple filling of states occupied by the exciton packet is enough to activate the effect, understanding the modifications of the dielectric properties of the material is more involved. We first consider the time-dependent change of the
static dielectric function for momenta spanning the BZ [Fig.~\ref{fig:screening}(a)]. The increase of the dielectric function is the key signature of \oscreen.

Comparing the evolution of the static dielectric constant [see different lines in Fig.~\ref{fig:screening}(a)] and of the distribution of carriers in $\mathbf{k}$-space [Fig.~\ref{fig:carriers}\,(c)] suggests that the  \oscreen\, effect follows the same trend as the
\paulib: the screening still increases for later times when the carriers scatter into the \ears. We note that the small momentum component $\vec{q}=0$ remains unchanged
at all times because the long range interaction is not screened due to the reduced dimensionality of the crystal~\cite{Latini2015}.

The rationale behind the trend of the  \oscreen\, can be captured by inspecting the independent particle dielectric function~\cite{Sangalli2017}, which has a simple dependency on the excited state occupations and energies:
\begin{equation}
\epsilon^{IP}_s (\mathbf{q}) \propto 1-\frac{8\pi}{q^2 V}\sum_{e,h,\kk} \frac{\Delta f_{\kk,\mathbf{q}}}{\Delta E_{\kk,\mathbf{q}}} \lvert
D_{eh}(\vec{k},\vec{q})\rvert^2,
\end{equation}
where $e,h$ and $\kk$ sum over the electrons, holes and momenta, $\Delta f_{\kk,\mathbf{q}}=f_{e,\mathbf{k}-\mathbf{q}}-f_{h\mathbf{k}}$ are the electron--hole
occupations, $\Delta E_{\kk,\mathbf{q}}=E_{e,\mathbf{k}-\mathbf{q}}-E_{h\mathbf{k}}$ their energies and $D_{eh}(\vec{k},\vec{q})$ is the dipole matrix element, describing the coupling of the electron--hole pairs with the external electric field.

Clearly, states with a small momentum transfer $\vec{q}$ will be weighted more strongly due to the $1/q^2$ prefactor, but also states with a small energy difference
$\Delta E_{\kk,\mathbf{q}}$ will
yield a larger contribution. Consequently, as the carriers relax to the direct band gap, i.e., to the \ears, they contribute more strongly to the
screening of the Coulomb interaction. This behavior is also reflected in the modification of the energy levels and the exciton binding energy, where the
screening directly enters via the statically screened Coulomb interaction $W(\vec{q},\tau)=V_\mathrm{C}(\vec{q})/\varepsilon_s(\vec{q},\tau)$, with
$V_\mathrm{C}$ the bare Coulomb potential. The evolution of
the average band gap around $\mathrm{K}$ for different pump energies [Fig.~\ref{fig:screening}(b)] reveals a two--stage behavior: an instantaneous resonant buildup of
the band gap change while
the pump is still active ($\tau=20$\,fs) whose efficiency decreases with the excess energy above the A-exciton. In the second phase, the strength of the BGR
slowly approaches the value obtained when the A-exciton is pumped resonantly. This part of the dynamics is governed by the relaxation of excited carriers to the
\ears where the  \oscreen\, effect is most pronounced.

\begin{figure}[H]
\includegraphics[width=3.25in]{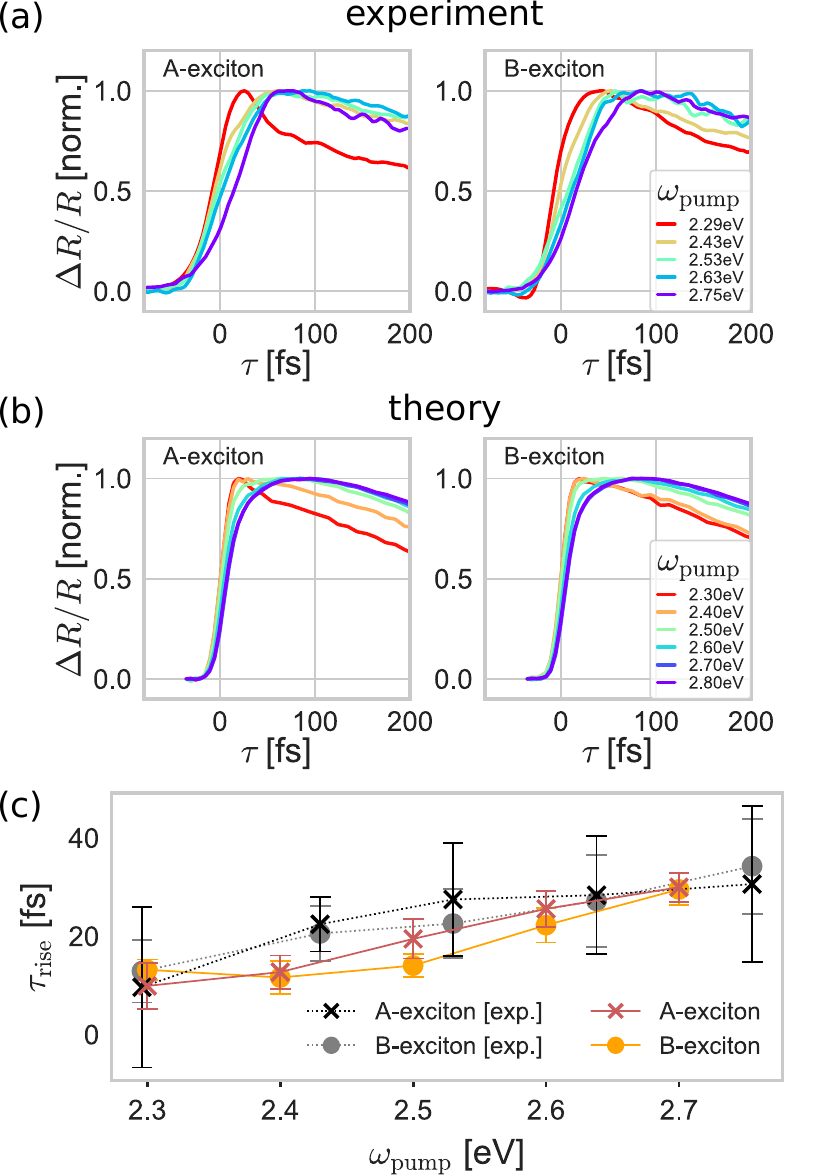}
\caption{Theoretical\,(frames b) time-dependent, constant energy cuts of the transient reflection signal at the A- and B- excitonic resonance compared to the experimental results of Ref.~\inlinecite{Trovatello2020g} (frames a). (c) Extracted rise time of the signal. Colored symbols: data from this work. Black and grey symbols: experimental data.}
\label{fig:final_evaluation}
\end{figure}

We next compare the time-dependent constant energy cuts of the transient reflection through the A- and B-excitonic resonance with the experimental results [Fig.~\ref{fig:final_evaluation}(a--b)] and find excellent agreement. The absorption onset rise time increases with the pump energy, $\omega_\mathrm{pump}$, and quickly saturates. The calculations also reproduce the different evolution of the saturation when the pump is resonant with the A-/B-exciton. Compared to the B-exciton case,
the A-exciton signal rises more rapidly with an overshoot for $\omega_\mathrm{pump}=2.3$\,eV which is absent for $\omega_\mathrm{pump}=2.4$\,eV, matching the experimental results.

To better highlight the quantitative agreement between theory and experiment, we fit the buildup signals with a convolution of the pump pulse (FWHM=20\,fs)
with an exponential rise and an exponential decay and extract the rise time of the signal $\tau_\mathrm{rise}$. Comparison with the measured values again shows remarkable agreement [Fig.~\ref{fig:final_evaluation}(c)]. Theory and experiment obtain the same
values and also the same dependence on the pump energy, $\omega_\mathrm{pump}$: 
a continuous increase of $\tau_\mathrm{rise}$ from $<10~\mathrm{fs}$ to $\tau_\mathrm{rise}\approx 30~\mathrm{fs}$. The present results even reproduce the 
 slight splitting observed in Ref.~\inlinecite{Trovatello2020g} in the rise times of the A- and B-exciton.

\begin{figure}[H]
	\includegraphics[width=3.25in]{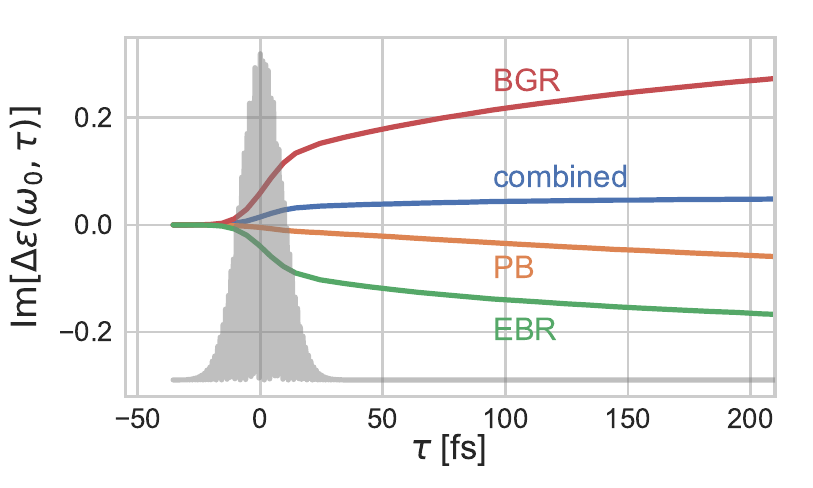}
	\caption{Decomposition of the different contributions to the change of the optical absorption $\Delta A(\omega_0,\tau)\propto\Delta\mathrm{Im}\left[\varepsilon(\omega_0,\tau)\right]$ for a pump energy of $\omega_\mathrm{pump}=2.50$\,eV. The traces are taken at an energy $\omega_0=1.87$\,eV slightly below the maximum of the A-excitonic peak. The grey shaded area is the time-profile of the pump pulse.}
	\label{fig:TR_no_scattering}
\end{figure}

We can now disentangle the relative importance of the different contributions to the evolution of the transient reflection spectra, namely the BGR, the EBR
and PB. For this, we consider the different ways how the excitation changes the optical absorption $\Delta
A(\omega_0,\tau)\propto\Delta\mathrm{Im}\left[\varepsilon(\omega_0,\tau)\right]$, see Eq.\,\ref{equ:BSE} in the Methods section. We evaluate the change at an
energy $\omega_0$ slightly below the A-excitonic peak, so that an increase (decrease) stands for photoinduced absorption (bleaching). 
In accordance with previous investigations of monolayer MoS$_2$, we find that the BGR governs the change of the optical response induced by the
pump~\cite{Pogna2016,Wood2020}, Fig.~\ref{fig:TR_no_scattering}: while EBR and PB alone would cause a bleaching of the signal, the full signal, i.e., the
combination of all effects, is dominated by a photoinduced absorption caused by BGR. 
We clearly see a two-stage dynamics connected to  \oscreen\, in the EBR and BGR dynamics, a signal buildup with the pump field followed by a slower rise caused by
the electron-phonon driven relaxation. This two-stage process is absent in the PB, which is a direct effect of carriers scattering into the \ears and shows on a
longer time scale. 

\section{CONCLUSIONS}

We investigate the ultrafast buildup of the transient reflection signal of monolayer MoS$_2$ on a SiO$_2$/Si substrate. The simulations are based on non--equilibrium Green's functions whose equations of motion are solved in a Kohn--Sham basis, where electrons and atoms are described fully \emph{ab--initio}, without adjustable parameters. Our approach is predictive and describes the carrier induced change of the screened electron-electron and electron-hole interactions, the phonon mediated scattering, and the effect of the substrate. We successfully interpret the pump-energy-dependent increase in the rise time of the signal as due to carrier migration to \ears close to the K points at the corners of the hexagonal BZ of the crystal. 
The delayed excess-energy-dependent  \oscreen\, by carriers scattering into those
regions leads to a delay in the rise time of the signal. This mechanism is confirmed by excellent agreement with the available experimental data, proving that the experimental observation of a delayed rise time is not a unique signature of an exciton formation but is rather dominated by the time-dependence of the band gap renormalization.


\section{METHODS}
{\bf Computational Details}.
The simulations are based on the Density--Functional Theory (DFT) ground state of a monolayer of MoS$_2$ in the local-density approximation (LDA)
obtained using the Quantum Espresso package~\cite{Giannozzi2020}. We use the bulk experimental value of
$5.97$\,Bohr ($3.16$\,\AA)~\cite{Dickinson1923} for the lattice constant $a_0$ and introduce an interlayer-spacing of $25$\,Bohr ($13$~\AA)~to avoid spurious interactions between the layers. We include two valence and 8 conduction bands on a $30\times30\times1$ Monkhorst-Pack grid and employ a denser incommensurable $61\times61\times1$
double-grid~\cite{Kammerlander2012} to obtain the $\vec{k}$-dependent lifetimes~\cite{Sangalli2019b}. To account for the 2D nature of the material, we
employ a cutoff in the Coulomb potential~\cite{Rozzi2006}.

The pump electric field is a Gaussian envelope sinusoidal pulse with different carrier frequencies $\omega_\mathrm{pump}$, an intensity full-width half-maximum of
$20~\mathrm{fs}$ and a peak intensity of $5\cdot 10^4~\mathrm{W/cm}^2$ inside the material.

The change of the eigenenergies of the system due to the excited carriers is calculated in the Coulomb hole and screened exchange (COHSEX) approximation. The
time-dependent optical properties of the material are then simulated by inserting the time-dependent change of the bands
$\epsilon^{QP}_{n\vec{k}}(\tau)=\epsilon^0_{n\vec{k}}+\Delta \epsilon^\mathrm{COHSEX}_{n\vec{k}}$, the time-dependent carriers $f_{n\vec{k}}(\tau)$ and the
screened Coulomb interaction $W(\vec{q},\tau)=V_\mathrm{C}(\vec{q})/\varepsilon_s(\vec{q},\tau)$ into the kernel of the Bethe-Salpeter equation
\begin{equation}
H^{exc}_{vc{\bf k}\\v'c'\bf{k}'} = \underbrace{\left( \epsilon^{QP}_{c{\bf k}}(\tau)-\epsilon^{QP}_{v{\bf k}}(\tau)\right)}_{\mathrm{BGR}}
\delta_{c,c'}\delta_{v,v'}\delta_{\bf k,k'} + \underbrace{\left( f_{c{\bf k}}(\tau)-f_{v{\bf k}}(\tau)\right)}_{\mathrm{PB}}\left[2\bar{V}_{vc{\bf
k}\\v'c'\bf{k}'} - \underbrace{W_{vc{\bf k}\\v'c'\bf{k}'}(\tau)}_{\mathrm{EBR}} \right]~~~,
\label{equ:BSE}
\end{equation}
where $v,c$ stand for the valence and conduction band and $\bar{V}$ is the bare Coulomb interaction without the long-range component.
\clearpage
\newpage
\begin{acknowledgement}
We thank Chiara Trovatello, Giulio Cerullo and Stefano Dal Conte for sharing the experimental data of Ref.~\inlinecite{Trovatello2020g}. A.M. acknowledges the funding received from the European Union projects: MaX {\em Materials design at the eXascale} H2020-EINFRA-2015-1, Grant agreement n.~676598, and H2020-INFRAEDI-2018-2020/H2020-INFRAEDI-2018-1, Grant agreement n.~824143; {\em Nanoscience Foundries and Fine Analysis - Europe} H2020-INFRAIA-2014-2015, Grant agreement n.~654360. V.S. acknowledges financial support by the doctoral college program ``TU-D Unravelling advanced 2D materials'' funded by TU Wien and the IMPRS-APS program of the MPQ (Germany). A.M.-S. acknowledges the Ramón y Cajal programme (grant RYC2018-024024-I; MINECO, Spain). L. W. acknowledges funding by the FNR (Fond National de Recherche, Luxembourg) via project INTER/19/ANR/13376969/ACCEPT.
\end{acknowledgement}

\begin{suppinfo}
Computational and theoretical details.
\end{suppinfo}

\bibliography{bibliography}

\end{document}